\documentclass[aps,prl,twocolumn,groupedaddress,showpacs]{revtex4}
\usepackage{epsfig}

\def\q{{\bf q}}
\def\p{{\bf p}}

\def\8{\infty}
\def\oh{\frac{1}{2}}

\def\d{\partial}

\def\undertext#1{\vtop{\hbox{#1}\kern 1pt \hrule}}

\def\be{\begin{equation}}
\def\ee{\end{equation}}
\def\bea{\begin{eqnarray} & &}
\def\eea{\end{eqnarray}}

\def\rf#1{(\ref{#1})}

\def\rf#1{(\ref{#1})}

\def\rfs#1{Eq.~\rf{#1}}

\begin{document}


\title{Strongly-resonant $p$-wave superfluids}


\author{J. Levinsen$^{1,2}$, N. Cooper$^{1,3}$ and V. Gurarie$^{1,2}$}
\affiliation{$^1$Institute for Theoretical Physics, University of California Santa Barbara, Santa Barbara CA 93106, USA\\
$^2$Department of Physics, University of Colorado,
Boulder CO 80309, USA \\
$^3$T.C.M. Group, Cavendish Laboratory, J. J. Thomson Avenue, Cambridge, CB3 0HE, United Kingdom}


\date{\today}

\begin{abstract}
  We study theoretically a dilute gas of identical fermions interacting via a
  $p$-wave resonance. We show that, depending on the microscopic physics,
  there are two distinct regimes of $p$-wave resonant superfluids, which we
  term ``weak'' and ``strong''.  Although expected naively to form a BCS-BEC
  superfluid, a strongly-resonant $p$-wave superfluid is in fact unstable
  towards the formation of a gas of fermionic triplets. We examine this
  instability and estimate the lifetime of the $p$-wave molecules due to the
  collisional relaxation into triplets. We discuss consequences for the
  experimental achievement of $p$-wave superfluids in both weakly- and
  strongly-resonant regimes.

\end{abstract}

\maketitle

Recently there has been considerable interest in trying to create a $p$-wave
resonant superfluid experimentally. The BCS and BEC regimes for such
superfluids are not just different aspects of the same phase, as they are for
the $s$-wave resonant superfluids, but rather are different phases. Thus the
tuning from the BCS to BEC regime involves a phase transition (or sometimes a
sequence of phase transitions) \cite{Gurarie2005,Yip2005}. Such a transition
can even be topological in some cases \cite{Read2000,
  Volovik2004,VolovikBook1,VolovikReview,Gurarie2005}. If the superfluid
is confined to two dimensions, the BCS phase will be topological and will
support vortices with non-Abelian excitations \cite{Read2000,Gurarie2005}.

In this paper we show that resonant $p$-wave superfluids must be classified as
two distinct types, with weak or strong Feshbach resonances (to be defined
precisely later).  The existing mean field theory of $p$-wave superfluids,
worked out in \cite{Gurarie2005,Yip2005,Gurarie2007}, applies only to the case
of weak Feshbach resonances. However, as we shall establish below, the
$p$-wave resonance used in ongoing experiments on
$^{40}K$\cite{Ticknor2004,Gaebler2007} is a strong resonance. It is therefore
important to determine the properties of strongly-resonant $p$-wave
superfluids.

The full theory of strong $p$-wave resonances is yet to be constructed.
Here we investigate an effect first noticed by Y. Castin and collaborators
\cite{CastinUnpublished}: in the regime of strong $p$-wave resonances the
fermions form triplet states with angular momentum (spin) 1. Superficially
similar to Efimov states \cite{Braaten2007}, these triplets are quite unusual.
They are very strongly bound, with a binding energy largely independent of
detuning from the resonance, as long as the detuning is not too large (but
dependent on the strength of the resonance). Correspondingly, their size is of
the order of the closed channel bound molecular state, far smaller than the
average interparticle separation. We find the critical value of the
resonance's strength at which the triplets first appear, and calculate their
binding energy as a function of the resonance strength.

Thus if a BEC condensate of strongly-resonant $p$-wave molecules is created,
one of its main channels of decay will be by molecular inelastic collisions,
with two molecules turning into one atom and one triplet. We estimate the
molecular lifetime due to this process and compare this with experimental
observations\cite{Gaebler2007}. We discuss limitations on the achievement
of $p$-wave superfluids in both weak and strong resonances, arising from this and
other inelastic decay processes.


The theory developed here can be used to investigate the true ground state of
a strongly-resonant $p$-wave condensate. This is likely to be a gas of
fermionic spin 1 triplets (or possibly of larger composite particles).

We consider a $p$-wave resonantly coupled superfluid, whose Hamiltonian is given by~\cite{Gurarie2005,Yip2005,Gurarie2007}
\begin{eqnarray}
\label{eq:ham} H &=& \sum_{p} \frac{p^2}{2m} ~\hat a^\dagger_{\bf p}
\hat a_{\bf p} + \sum_{{\bf q}, \mu} \left(\epsilon_0 +
{\frac{q^2}{4m}} \right)
\hat b_{\mu{\bf q}}^\dagger \hat b_{\mu {\bf q}}\\
&+&\sum_{{\bf p},{\bf q},\mu} ~{g ( \left| \p \right|) \over \sqrt{V}} \left( \hat b_{\mu
{\bf q}} ~p_\mu~ \hat a^\dagger_{{\q\over 2}+\p} ~\hat a^\dagger_{{\q
\over 2}-\p} + h. c. \right).\nonumber
\end{eqnarray}
Here $\hat a^\dagger$, $\hat a$ are the creation and annihilation operators of a spinless fermion (atom) with mass $m$, and $\hat b^\dagger_\mu$, $\hat b_\mu$
are the creation and annihilation operators of a bosonic diatomic molecule of
spin 1 (the 3D vector index $\mu$ represents the projection of spin).
This superfluid is controlled by four parameters. The first two  are the detuning $\epsilon_0$ and the overall particle number $N$, an expectation value of the operator
$ \hat N = \sum_p \hat a^\dagger_p \hat a_p + 2 \sum_{\mu, {\bf q}} \hat b^\dagger_{\mu \bf q} \hat b_{\mu \bf q}.$
It is more convenient to work with the energy equivalent of $N$, the Fermi energy
$ \epsilon_F = \left( 6 \pi^2 \hbar^3 N /V\right)^{2/3 }/(2m).$ The other two are contained in the coupling constant $g(\left| \p \right| )$.
The physical origin of the dependence of $g$ on $\left| \p \right|$ lies in the fact that the molecules have finite size. This can be captured by choosing
$g$ to remain constant as long as $|\p| \ll \Lambda$ (which we denote simply by $g$) and quickly drop to zero if $|\p| \gg \Lambda$.
Here $R_e \sim \hbar/\Lambda$  is the physical (closed-channel) size of the molecules.
The knowledge of exactly how $g$ drops to zero at large momenta may be
important. In this paper we adopt the ``hard momentum cutoff'' approach
$g(|\p|) = g\, \Theta(\Lambda-p)$ ($\Theta$ is equal to $1$ or $0$ depending
on whether its argument is positive or negative). We have studied other types
of cutoff, and find these do not change the main conclusions of this paper.

Two dimensionless parameters can be constructed out of $g$, $\epsilon_F$,  and $\Lambda$, namely
\begin{equation} \label{eq:gamma}
\gamma = \frac{m^{\frac 5 2} g^2 \sqrt{\epsilon_F}}{\hbar^3}, \ c_2 = \frac{m^2 g^2 \Lambda}{3\pi^2 \hbar^3}. \end{equation}
Notice that in order to observe universal (short distance physics independent) behavior, the interparticle separation ($\sim \hbar /\sqrt{m\epsilon_F}$) must be kept much bigger than $R_e$, thus $\gamma \ll c_2$.

Both of these parameters control the perturbative expansion of \rf{eq:ham} in powers of the coupling $g$. It is customary, when analyzing \rfs{eq:ham}, to
apply a mean field approximation whose validity is based on the smallness of $g$. Strictly speaking, both $\gamma$ and $c_2$ must be small
in order for the mean field approximation employed in the original  publications
investigating \rfs{eq:ham}~\cite{Gurarie2005,Yip2005} to be valid. $\gamma$ depends on the interparticle separation and can be
made small simply by reducing the particle density. $c_2$ however depends
solely on the physics of the Feshbach resonance which led to \rfs{eq:ham}; its
value, which can be small or large, is fixed by the atomic type and Feshbach
resonance involved, so it cannot be continuously controlled.

One terms the superfluids with $\gamma \ll 1$ as those with {\it narrow} Feshbach resonances, while the ones with $\gamma \gg 1$ are the {\it broad} Feshbach resonance superfluids \cite{Andreev2004,Gurarie2005,Gurarie2007}.
Likewise, we will term the $c_2 \gg 1$ resonances as the {\it strong} $p$-wave Feshbach resonances, while those with $c_2 \ll 1$ are {\it weak} resonances.
The $p$-wave resonances are typically narrow because, even if they are not,
they can be made narrow by reducing the particle density.

The narrow and weak $p$-wave resonances have been thoroughly investigated in prior publications. It is therefore imperative to consider the narrow and strong  resonances. The main idea behind the analysis is based on the fact that fluctuational corrections to the mean field come from two distinct regions in momentum space, $p$ of order $\hbar/l$ where $l$
is interparticle spacing, and $p$ of order $\Lambda$. The former capture the many-body physics of \rfs{eq:ham} and are small as long as $\gamma$ is small. The latter
come from high momenta and energies at which
no real particles propagate.
Thus this contribution, controlled by $c_2$, is essentially few-body,
equivalent to solving some few-body Schr\"odinger equation, which although difficult, is not an impossible task.


The analysis carried out in Ref.~\cite{Gurarie2007} showed
that the $p$-wave Feshbach resonance in $^{40}$K used in Refs.~\cite{Ticknor2004,Gaebler2007} was strong.  The derivation relied on the scattering amplitude of two atoms
calculated in Ref.~\cite{Gurarie2005,Gurarie2007}
\begin{equation} \label{eq:genpscatamp}
f(k) = \frac{k^2}{-\frac{1}{v}+\oh k_0 k^2 - i k^3}.
\end{equation}
where
\begin{equation} \label{eq:par}
v=- \frac{m g^2}{6 \pi \hbar \left(1+c_2 \right) \omega_0}, \ k_0 =
- \frac{ 4 \Lambda \left(1 + c_2 \right)}{\pi c_2}.
\end{equation}
Here $v$ is the so-called effective volume, controlled by the physical detuning $\omega_0$ [related to $\epsilon_0$ by $\omega_0 =( \epsilon_0 -m g^2 \Lambda^3/({9 \pi^2 \hbar^2}))/{(1+c_2)}$]
and $k_0$ is a parameter similar to the effective range of $s$-wave scattering (having, however, the
dimensions of inverse length). If $k_0$ and $\Lambda$ are known (numerically or experimentally), $c_2$ can be found from \rf{eq:par}.

We also remark that had we considered a one-channel model of identical fermions interacting via a short range $p$-wave potential, such as $V(r) =\lambda \, \d_\mu \delta^{(3)}(r) \d_\mu$, we would have obtained \rfs{eq:genpscatamp} with $k_0 \sim -\Lambda$ \cite{Gurarie2007}. In other words, such a model automatically describes strong resonances.

\begin{figure}[bt]
\includegraphics[height=.35in]{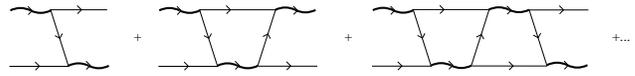}
\caption{The diagrams whose sum gives the scattering amplitude between an atom and a molecule}
\label{Fig-Feshbach}
\end{figure}

We now turn our attention to the physical consequences of strong resonances. Its main consequence is the existence of a bound state of three atoms when $c_2$ exceeds a certain threshold.  To show this, we calculate the scattering amplitude of one atom and one molecule. This is given by a sequence of diagrams depicted on Fig.~\ref{Fig-Feshbach}. These diagrams are identical to the ones studied in the context of the $s$-wave BCS-BEC crossover \cite{Brodsky2005,Levinsen2006}.

Here the atoms propagate with the free propagator $G(\p,\omega) = 1/ \left({\omega-\frac{p^2}{2m}+i0} \right)$, while to find the molecular propagator one needs to calculate its self-energy
~\cite{Gurarie2007}
\begin{eqnarray}
&&
D_{\mu \nu}(\q,\omega) = \delta_{\mu \nu} / \left[ (1+c_2)\left(
\omega-\frac{q^2}{4m} -\omega_0 +i0\right) \right.  \nonumber \\ &&  \hspace{-5mm}\left. +c_2 \frac{\sqrt{m}}{\Lambda} \left( \frac{q^2}{4m} - \omega -i0 \right)^\frac 3 2 \arctan \left( \frac{\Lambda} {\sqrt{q^2/4 - m \omega }} \right) \right]
\end{eqnarray}
(In these, and subsequent, expressions we set $\hbar=1$ for clarity.)
Each loop in the diagrams on Fig.~\ref{Fig-Feshbach} is linearly divergent. It is this divergence, occurring at momenta $p \sim \Lambda$ and controlled by $c_2$, which we would like to capture. To do so,
we study the atom-dimer scattering problem with the following
kinematics; a boson of spin $\mu$ and 4-momentum $({\bf
  0},\kappa+E_3)$ scatters off a fermion with 4-momentum $({\bf
  0},0)$. The outgoing particles are a boson with spin $\nu$ and
4-momentum $(\q,q_0+\kappa+E_3)$ and a fermion with $(-\q,-q_0)$. Here
$\kappa(\omega_0)$ is an implicit function of the detuning such that
the bosonic propagator $D(\q,q_0+\kappa)$ has a pole as
$\q,q_0\to0$. $E_3\leq0$ is the energy at which we are looking for a
bound state. The scattering $T$-matrix has the following general form
\begin{equation}
T_{\mu\nu}(\p,p_0) = T_1(p,p_0)\delta_{\mu\nu}+T_2(p,p_0)p_\mu p_\nu/p^2
\end{equation}
and the scattering length $a_{\rm bf}$ is related to $T_1(0,0)$
(evaluated at $E_3=0$) as
$a_{\rm bf} = \frac m{3\pi}T_1(0,0)$.

The integral equation for the $T$-matrix is derived analogously to the
$s$-wave problem \cite{Brodsky2005,Levinsen2006}, and is
\begin{widetext}
\begin{eqnarray}
T_{\mu\nu}(\p,p_0) & = & -\frac{2}{1+c_2}G(\p,p_0+\kappa+E_3)
p_\mu p_\nu g(|\p|)g(|\p|/2)
-4i\int\frac{d^4q}{(2\pi)^4}T_{\mu\alpha}(\q,q_0)
D(\q,q_0+\kappa+E_3)G(-\q,-q_0)
\nonumber \\ && \hspace{5mm}\times
G(\p+\q,p_0+q_0+\kappa+E_3)(p+q/2)_\alpha(q+p/2)_\nu
g(|\p+\q/2|)g(|\q+\p/2|)
\end{eqnarray}
\end{widetext}
The factor $1+c_2$ is the inverse residue of
the bosonic propagator. $T_{\mu\nu}(\vec q,q_0)$ is analytic in the
upper halfplane of $q_0$ and thus we may integrate out $q_0$, setting
$q_0\to-q^2/2m$. To solve the integral equation we then let $p_0\to
-p^2/2m$. For simplicity define $T_i(p,-p^2/2)\equiv
T_i(p)$. Measuring momenta in units of the cutoff, energies in units
of $\Lambda^2/m$, and the $T$-matrix itself in units of $1/(m\Lambda)$
we find the integral equation
\begin{eqnarray}
T_j(p) & = & 6\pi^2\frac{c_2}{1+c_2}\frac{p^2}
{p^2-\kappa-E_3}\Theta(1-p)\delta_{2j} \nonumber \\ && \hspace{-20mm}
-3c_2\int_0^2q^2dq\, D(q,-q^2/2+\kappa+E_3)a_{ji}(p,q)T_i(q)
\label{eq:t3}
\end{eqnarray}
The coefficients $a_{ji}(p,q)$ are given by an integration over
directions of $\q$
\begin{eqnarray}
a_{ji}(p,q) & = &
\int\frac{d\Omega_{\q}}{4\pi}\frac{\left(\begin{array}{cc} 1 & -1\\
-1 & 3\end{array}\right)_{jk}\left(\begin{array}{c}\delta_{\mu\nu}\\ \frac{p_\mu p_\nu}{p^2}
\end{array}\right)_k\left(\begin{array}{c}\delta_{\mu\nu}\\\frac{q_\mu q_\alpha}{q^2}
\end{array}\right)_i}{\kappa+E_3-p^2-q^2-\p\cdot \q+i0}
\nonumber \\ &&
\hspace{-17mm}\times \frac1{g^2}(p+q/2)_\alpha(q+p/2)_\nu g(|\p+\q/2|)g(|\q+\p/2|)
\label{eq:aij}
\end{eqnarray}

\begin{figure}[bt]
\includegraphics[height=2.6in,angle=270]{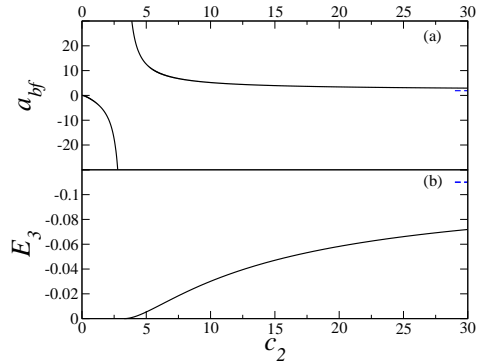}
\caption{(a) Scattering length $a_{\rm bf}$ in units of $\hbar/\Lambda$
  and (b) binding energy $E_3$ of the triplet in units of
  $\Lambda^2/m$, both as functions of $c_2$. Here, detuning has been
  set to zero. The large $c_2$ limit is indicated.}
\label{fig:abfe3}
\end{figure}

The scattering length $a_{\rm bf}$ is found by solving
Eq.~(\ref{eq:t3}) at $E_3=0$. The binding energy
of the triplet corresponds to a pole in the $T$-matrix and thus to a
solution of the homogeneous integral equation at a specific value of
$E_3$.  Fig.~\ref{fig:abfe3}(a) shows how the scattering length
is negative for a weak Feshbach resonance, becoming
more negative and diverging at $c_2\approx3.3$. This is the strength
of the resonance at which the bound triplet appears, as illustrated in
Fig. \ref{fig:abfe3}(b). As $c_2\to \infty$ the scattering length saturates
at $a_{\rm bf}\approx1.9/\Lambda$ and the binding energy at
$E_3\approx-0.11\Lambda^2/m$. The existence of a finite $c_2 \rightarrow \infty$ limit can also be
 seen by observing that each of the diagrams depicted on Fig.~\ref{Fig-Feshbach} has a finite
 $c_2 \rightarrow \infty$ limit.
 It should be noted that whereas the
presence of the bound state and the general features of the scattering
length and binding energy do not depend on the method of cutoff,
the exact numerical values will in general depend on the method
chosen.

The existence of the bound trimer state for large $c_2$ raises the possibility
of an inelastic decay channel in which two dimers collide to leave a trimer
and an unbound atom (with large relative velocity). (Henceforth we use the
term ``dimer'' to refer to a molecule of two atoms, to distinguish this
clearly from a triplet, or trimer.)  In a non-degenerate gas of dimers, these
inelastic losses will cause the density of dimers $n_{\rm d}$ to decay as
\begin{equation}
\frac{d n_{\rm d}}{dt} = - \alpha_{\rm dd} n_{\rm d}^2
\end{equation}
with $\alpha_{\rm dd} = 2 \frac{\hbar}{m} \langle k_i \sigma_{\rm
  in}(k_i)\rangle$ where the average is over the relative momenta of the
incident dimers, $k_i$, and $\sigma_{\rm in}(k_i) = \int |f_{\rm in}|^2
d\Omega \frac{k_f}{k_i}$ with $f_{\rm in}$ the inelastic scattering amplitude
into a final momentum $k_f$. Arguments similar to the ones presented above
for dimer-atom scattering show that
 $|f_{\rm in}|^2\sim R_e^{2}$.  Thus, for
large $c_2$, such that the trimer
binding energy is $-E_3 \sim \hbar^2/(mR_e^2)$ and is large compared to the incident kinetic energy, one
finds
\begin{equation}
\label{eq:estimate}
\alpha_{\rm dd}
{\sim}
 \frac{\hbar}{m}R_e \,.
\end{equation}

It is instructive to compare this result with the inelastic decay constants
into deep bound states for $s$-wave dimers, formed from (two-component)
fermions or from bosons with $s$-wave scattering length $a$. Close to the
$s$-wave resonance, $a \gg R_e$, the decay constant (\ref{eq:estimate}) is
much smaller than that expected for bosons, $\alpha^{s-{\rm boson}}_{\rm dd}
\sim {\hbar a}/{m}$, but is larger than that for $s$-wave dimers of fermions,
$\alpha^{s-{\rm fermion}}_{\rm dd} \sim \frac{\hbar R_e}{m}
(R_e/a)^{2.55}$\cite{Petrov2005}.  The suppressed decay of $s$-wave dimers of
fermions is explained in Ref.\cite{Petrov2005} as an effect of the Pauli
principle, reducing the probability to find three atoms within a lengthscale
$R_e$.  In a $p$-wave dimer the two atoms have a probability of order unity to
be inside the centrifugal barrier, at a separation of order $R_e$. Taking this
feature of the $p$-wave dimers into account, simple estimates lead to
$\alpha_{\rm dd} \sim \hbar R_e/m$ for decay into trimers, consistent with the
result (\ref{eq:estimate}) from the $T$-matrix calculation.  In addition to
this channel, there are inelastic channels -- active for both weak and strong
resonances -- involving decay into deep dimer states.  Applying the same
simple estimates, one finds that the inelastic decay constants for dimer-dimer
and dimer-atom scattering are also $\alpha_{\rm dd}\sim \alpha_{\rm da} \sim
\hbar R_e/m$.

In recent experimental work \cite{Gaebler2007} a gas of $p$-wave Feshbach
dimers was created in $^{40}$K. Unfortunately the lifetime of the dimers was
observed to be quite short, about 2 ms. While $^{40}$K can suffer losses
through dipolar relaxation (an effect expected to be absent for $p$-wave
resonances in other fermionic systems, for example $^6$Li),
Ref.~\cite{Gaebler2007} found that the lifetime was shorter than that
predicted for dipolar relaxation alone.  Additional losses could arise from
inelastic collisions of the dimers. This mechanism would imply a density
dependence of the decay rate; this dependence has not, as yet, been
established experimentally.

Within the above considerations, we expect the decay rate of dimers via
relaxation into deep trimers or dimers under inelastic collisions (with other
dimers or with unbound atoms) to be of order $\Gamma_{\rm in} \sim \frac{\hbar
  R_e}{m} n$, where $n$ is the density of atoms or dimers with which a given
dimer can collide.  Taking $n \simeq 7 \times 10^{12}\mbox{cm}^{-3}$ (the
atomic density in the experiments of Ref.\cite{Gaebler2007}) we find
$\Gamma_{\rm in} \sim 10 \mbox{Hz}$. This estimate is more than one order of
magnitude smaller than the additional decay rate required to account for the
observations of Ref.~\cite{Gaebler2007}.  However, we note that the prefactor
to the estimate is uncertain. In view of this uncertainty, and in view of the
lack of clear evidence of a density dependence in the experiment, it remains
an open issue whether the dimer lifetime in Ref.\cite{Gaebler2007} is limited
by inelastic collisions.


Our analysis has important consequences for possibilities to achieve
superfluid phases close to a $p$-wave resonance.  On the BEC side of the
resonance, our calculations show that the elastic dimer-dimer scattering
amplitude is $f_{\rm el} \sim R_e$. Consequently, the elastic scattering rate
is of order $\Gamma_{\rm el} \sim \frac{\hbar R_e}{m}{n}_{\rm d} ({k_i R_e})$,
which is typically much smaller than the inelastic decay rate, $\Gamma_{\rm
  in} \sim \frac{\hbar R_e}{m}n_{\rm d}$.  (For a BEC of dimers, $k_i$ is
small compared to the inverse particle spacing, $1/l$, so $k_i R_e\lesssim
{R_e/l} \ll 1$.)  It is therefore unlikely that a BEC of dimers can undergo
sufficient elastic collisions to thermalize before inelastic losses deplete
the gas.  On the other hand, on the BCS side of the resonance, thermalization
can proceed at a much faster rate, and will be limited by the rate of
hybridisation of the dimers with the unbound atoms (this is the rate at which
pairs of atoms can exchange their relative momentum).  Assuming the densities
of dimers and atoms to be comparable, $ n_{\rm d}\sim n_{\rm a} \equiv n$, one
finds that the hybridisation rate, as set by the width of the resonance
(\ref{eq:genpscatamp}), is $\Gamma_{\rm hyb} \sim
\frac{\gamma}{1+c_2}\frac{\epsilon_F}{\hbar}$.  Using (\ref{eq:gamma}), we
find that $\Gamma_{\rm hyb}/\Gamma_{\rm in} \sim \frac{c_2}{1+c_2}$. Thus,
provided the resonance is not very weak ($c_2$ very small), the rate of
hybridisation is parametrically the same as $\Gamma_{\rm in}$, and the system may thermalize before inelastic
losses deplete the gas.  Thus our results show that {\it it is on the BCS side
  of a strong resonance that one has the best opportunity to attain a
  thermalized $p$-wave superfluid phase}.  Finally, we note that the
limitations we have described in this paragraph, arising from decay into deep
bound states, could be eliminated in an ``optical Feshbach'' scheme in which
the particles are coupled to a deep closed-channel molecule. In this case, it
is important that the resonance should be sufficiently weak in order also to
eliminate inelastic decay processes into the triplet states that always exist
for strong resonances.

The authors are grateful to the participants of the KITP, Santa Barbara,
program ``Strongly Correlated States in Condensed Matter and Atomic Physics"
 for many comments and useful discussions.
This work was supported by the NSF via grants DMR-0449521 and PHY-0551164,
and by EPSRC GR/S61263/01.

\bibliography{paper}

\end{document}